# Cyber Security Awareness Campaigns: Why do they fail to change behaviour?

Maria Bada[1], Angela M. Sasse[2] and Jason R.C. Nurse[3]

[1] Global Cyber Security Capacity Centre, University of Oxford maria.bada@cs.ox.ac.uk
[2] Department of Computer Science, University College London a.sasse@cs.ucl.ac.uk
[3] Cyber Security Centre, Department of Computer Science, University of Oxford
jason.nurse@cs.ox.ac.uk

## Abstract

The present paper focuses on Cyber Security Awareness Campaigns, and aims to identify key factors regarding security which may lead them to failing to appropriately change people's behaviour. Past and current efforts to improve information-security practices and promote a sustainable society have not had the desired impact. It is important therefore to critically reflect on the challenges involved in improving information-security behaviours for citizens, consumers and employees. In particular, our work considers these challenges from a Psychology perspective, as we believe that understanding how people perceive risks is critical to creating effective awareness campaigns. Changing behaviour requires more than providing information about risks and reactive behaviours – firstly, people must be able to understand and apply the advice, and secondly, they must be motivated and willing to do so – and the latter requires changes to attitudes and intentions. These antecedents of behaviour change are identified in several psychological models of behaviour. We review the suitability of persuasion techniques, including the widely used 'fear appeals'. From this range of literature, we extract essential components for an awareness campaign as well as factors which can lead to a campaign's success or failure. Finally, we present examples of existing awareness campaigns in different cultures (the UK and Africa) and reflect on these.

# 1   Introduction

Governments and commercial organizations around the globe make extensive use of Information and Communications Technologies (ICT), and as a result, their security is of utmost importance. To achieve this, they deploy technical security measures, and develop security policies that specify the 'correct' behaviour of employees, consumers and citizens. Unfortunately, many individuals do not comply with specified policies or expected behaviours [1]. There are many potential reasons for this, but two of the most compelling are that people are not aware of (or do not perceive) the risks or, they do not know (or fully understand) the 'correct' behaviour.

The primary purpose of cyber security-awareness campaigns is to influence the adoption of secure behaviour online. However, effective influencing requires more than simply informing people about what they should and should not do: they need, first of all, to accept that the information is relevant, secondly, understand how they ought to respond, and thirdly, be willing to do this in the face of many other demands [2][3].

This paper engages in a focused review of current literature and applying psychological theories to awareness and behaviour in the area of cyber security. Our aim is to take a first step towards a better understanding of the reasons why changing cyber security behaviour is such a challenge. The study also identifies many psychological theories of behavioural change that can be used to make information security awareness methods significantly more effective.

This paper is structured as followed. Section 2 reviews current information on security-awareness campaigns and their effectiveness. In Section 3, we examine the factors influencing change in online behaviour, such as personal, social and environmental factors. Section 4 reflects on persuasion techniques used to influence behaviour and encourage individuals to adopt better practices online. In Section 5, we summarise the essential components for a successful cyber security awareness campaign, and consequently, factors which can lead to a campaign's failure. Finally, Section 6, presents examples of existing awareness campaigns in the UK and Africa and initially reviews them in the light of our study's findings.

# 2   Cyber security awareness campaigns

An awareness and training program is crucial, in that, it is the vehicle for disseminating information that all users (employees, consumers and citizens, including managers) need. In the case of an Information Technology (IT) security program, it is the typical means used to communicate security requirements and appropriate behaviour. An awareness and training program can be effective, if the material is interesting, current and simple enough to be followed. Any presentation that 'feels' impersonal and too general as to apply to the intended audience, will be treated by users as just another obligatory session [4].

Security awareness is defined in NIST Special Publication 800-16 [4] as follows: *"Awareness is not training. The purpose of awareness presentations is simply to focus attention on security. Awareness presentations are intended to allow individuals to recognize IT security concerns and respond accordingly"*. This clearly highlights where the main emphasis on awareness should be. It identifies the fact that people need not only to be aware of possible cyber risks but also, behave accordingly.

In terms of the public more generally, governments encourage citizens to transact online and dispense advice on how to do so securely. However, major cyber security attacks continue to occur [5]. Although a likely reason for this could be the fact that attackers are becoming more skilled, there is also the reality that security interfaces are often too difficult for the layman to use.

Another relevant point that has arisen from the literature is the fact that people know the answer to awareness questions, but they do not act accordingly to their real life [6]. It is proposed that it is essential for security and privacy practices to be designed into a system from the very beginning. A system that is too difficult to use will eventually lead to users making mistakes and avoiding security altogether [7]. This was the case in 1999 [8] and is still the case today [9].

The fact today is that security awareness as conceived is not working. Naturally, an individual that is faced with so many ambiguous warnings and complicated advice, may be tempted to abandon all efforts for protection, and not worry about any danger. Threatening or intimidating security messages are not particularly effective, especially because they increase stress to such an extent that the individual may even be repulsed or deny the existence of the need for any security decision.

## 3 Factors influencing change in online behaviour

The increased availability of information has significant positive effects, but simply providing information often has surprisingly modest and sometimes unintended impacts when it attempts to change individuals' behaviour [10]. A considerable amount of investment is being spent by governments and companies on influencing behaviour online [11], and the success in doing so would be maximised if they draw on robust evidence of how people actually behave.

Various research articles have investigated the factors which influence human behaviour and behaviour change but one of the most complete is the Dolan, et al. [12]. In their article the authors present nine critical factors, namely: (1) the messenger (who communicates information); (2) incentives (our responses to incentives are shaped by predictable mental short cuts, such as strongly avoiding losses); (3) norms (how others strongly influence us); (4) defaults (we follow pre-set options); (5) salience (what is relevant to us usually draws our attention); (6) priming (our acts are often influenced by sub-conscious cues); (7) affect (emotional associations can powerfully shape our actions); (8) commitments (we seek to be consistent with our public promises, and reciprocate acts); (9) ego (we act in ways that make us feel better about ourselves).

These factors hint at the key ingredients for an overall approach to influencing behaviour change, since the psychological mechanisms which they refer to, are core in making any type of decision. Furthermore, these mechanisms can influence the user's motivation to actually adopt the knowledge offered by a security campaign and behave accordingly. In order to enact change, the current sources of influence (conscious or unconscious, personal, environmental or social) need to be identified. The following section describes these aspects.

### *3.1 Personal factors*

Reflecting on literature, it is well recognised that an individual's knowledge, skills and understanding of cyber security as well as their experiences, perceptions, attitudes and beliefs are the main influencers of their behaviour [13]. Of these, personal motivation and personal ability, are two of the most powerful sources of influence. Specifically, it is the difference between what people say and what people do that needs to be addressed. In many cases, people will have to overcome existing thought patterns in order to form new habits.

People can sometimes get tired of security procedures and processes, especially if they perceive security as an obstacle, preventing them from their primary task (e.g., being blocked from visiting a music download website because the browser has stated that the site might have malware). It can also be stressful to remain at a high level of vigilance and security awareness. These feelings describe the so called 'security fatigue', and they can be hazardous to the overall health of an organization or society [14][15].

In the security domain, the so called 'Security, Functionality and Usability Triangle', describes the situation of trying to create a balance between three, usually conflicting, goals [16]. If you start in the middle and move toward security, you also move further away from functionality and usability. Move the point toward usability, and you are moving away from security and functionality. If the triangle leans too far in either direction, then this can lead to a super secure system that no one can use, or an insecure system that everyone can use (even unwanted individuals, such as hackers). Security fatigue becomes an issue when the triangle swings too far to the security side and the requirements are too much for the users to handle. Therefore, there has to be a balance between system security and usability [9].

Moreover, perceived control is a core construct that can also be considered as an aspect of empowerment [17]. It refers to the amount of control that people feel they have, as opposed to the amount of their actual control [18][19][20]. The positive effects of perceived control mainly appear in situations where the individuals can improve their condition through their own efforts. Also, the greater the actual threat, the greater the value that perceived control can play. When we apply this theory to cyber security, we could assume that home-computer users often experience high levels of actual control over their risk exposure. This is because they can choose which websites to visit, whether to open email attachments and whether to apply system updates [21].

In Psychology, the Regulatory Focus theory [22] proposes that in a promotion-focused mode of self-regulation, individuals' behaviours are guided by a need for nurturance, the desire to bring oneself into alignment with one's ideal self ('ideal self' is what usually motivates individuals to change), and the striving to attain gains. In a prevention-focused mode of self-regulation individual's behaviours are guided by a need for security, the need to align one's actual self with one's ought self by fulfilling duties and obligations and the striving to ensure non-losses. Thus, the effectiveness of advertising campaigns for adolescents may be enhanced either by using two types of messages (prevention and promotion focused) or by priming one type of regulatory focus through the advertising vehicle.

### 3.2 Cultural and environmental factors

Culture is also an important factor that can have a positive security influence to the persuasion process. Messages and advertisements are usually preferred when they match the cultural theme of the message recipient. As a result, cultural factors are one of the most important factors for consideration when designing education and awareness messages [23].

The cultural systems of a society shape a variety of their psychological processes. Intrinsically motivated behaviours emanate from the self and are marked by the enjoyment and satisfaction of engaging in an activity. Conversely, extrinsic motivation refers to motivation to engage in an activity in order to achieve some instrumental end, such as earning a reward or avoiding a punishment. Messages tend to be more persuasive when there is a match between the recipient's cognitive, affective or motivational characteristics and the content of framing of the message. Also, messages are more persuasive if they match an individual's ought or self-guides, or self-monitoring style [24]. People might be motivated to follow a cyber security campaign's advice. But if that causes certain limitation on the sites they can visit online, then this can automatically result in emotional discomfort, thus leading to ignorance of a suggested 'secure' behaviour.

Perception of risk can be a collective phenomenon and it is crucial for awareness raising specialists to be aware of the different cultural characteristics. The values that distinguish country cultures from each other could be categorised into four groups [25]: (1) Power Distance; (2) Individualism versus Collectivism; (3) Masculinity versus Femininity; and (4) Uncertainty Avoidance. In more individualistic cultures, such as the West, people tend to define themselves in terms of their internal attributes such as goals, preferences and attitudes. For example, in cyber security, a message used in a Western country would tend to avoid presenting the general risks of not being secure online and rather focus on the benefits of being secure.

In more collectivist cultures, such as those typically found in the East, individuals tend to define themselves in terms of their relationships and social group memberships [26]. In this cultural context, individuals tend to avoid behaviours that cause social disruptions. Therefore, they favour prevention over promotion strategies focusing on the negative outcomes, which they hope to avoid rather than the positive outcomes they hope to approach [27]. Moreover, risk is also seen as the other side of trust and confidence, a perception being imbedded in social relations [28]. The emphasis on different risks, in different cultural contexts is another important aspect that needs to be addressed when creating cyber security awareness campaigns.

## 4  Persuasion techniques

Persuasion can be defined as the *"attempt to change attitudes or behaviors or both (without using coercion or deception)"* [29]. There are two ways of thinking about changing behaviour: (1) by influencing what people consciously think about (rational or cognitive model) and (2) by shaping behaviour focused on the more automatic processes of judgment and influence (context model) without changing the thinking. In this section we present the different persuasion and influence techniques, in an effort to examine potential challenges in the area of cyber security awareness.

### *4.1  Influence strategies*

People do not usually simply follow advice or instructions on how to behave online even if they come from an expert or a person of authority. In many cases, end users are not fully aware of the dangers of interacting online, and to exacerbate the issue, security experts provide them with too complicated information, often evoking emotions of fear and despair [30]. The basic persuasion techniques include: fear, humour, expertise, repetition, intensity, and scientific evidence.

People base their conscious decisions on whether they have the ability to do what is required and whether the effort is worth it. Examples of messages aimed at persuading individuals to change their behaviour online, can be found in advertising, public relations and advocacy. These 'persuaders' use a variety of techniques to seize attention, to establish credibility and trust, and to motivate action. These techniques are commonly referred to as the 'language of persuasion'. They can also be found in cyber security awareness campaigns. For example, fear is often being used as a persuasion technique for cyber security.

Surveys have shown that the invocation of fear can be a very persuasive tactic to specific situations, or indeed a counterproductive tactic in others [31]. Security-awareness campaigns mostly tend to use fear invocations, by combining messages with pictures of hackers in front of the screen of a computer. Even, the word 'cyberspace', indicates something unknown to many, thus leading to fear. Typically, invocations of fear, are accompanied with recommendations that are as efficacious in preventing the threat. Thus, the three central structures in fear invocations are fear, threat and efficacy.

Various behavioural theories including the Drive Model [32], the Parallel Response Model [33], or the Protection Motivation Theory [34], consider the cost and efficiency of a reaction and have independent effects on persuasion. According to the Protection Motivation Theory for instance, the way a person responds to and carries out a cyber security awareness campaign's recommendations depends on both the cyber threat appraisal but also on the person's self-efficacy.

The attempt to change a certain behaviour is much more difficult when the person is bombarded by a large number of messages about certain issues. However, even when the design of the message is taken into account, there is a big gap between the recognition of the threat and the manifestation of the desired behaviour at regular intervals. Specifically for security awareness campaigns, the behaviour that users will need to adopt, should be as simple and easy as possible highlighting the advantages of adopting it.

Moreover, findings suggest that interventions based on major theoretical knowledge to change behaviour (e.g., social learning theory or the theory of self-efficacy) that take into account cultural beliefs and attitudes, and are more likely to succeed [35].

# 5   Factors leading to success or failure of a cyber security awareness campaign

There are several components which need to be taken into consideration in order for an awareness campaign to be successful. One of the most crucial parts is that of *communication*. Teaching new skills effectively can lead to prevention of high-risk online behaviour, since what appears to be lack of motivation is sometimes really lack of ability [36].
There is a wide discussion about security-awareness campaigns and their effort to secure the human element, leading to a secure online behaviour. In many cases, security-awareness campaigns *demand a lot of effort and skills* from the public, while measures do not provide real insight on their success in changing behaviour. Often, solutions are not aligned to risks; neither progress nor value are measured; incorrect assumptions are made about people and their motivations; and unrealistic expectations are set [6].

As previously discussed *fear invocations* have often proved insufficient to change behaviour [31]. For example, a message combined to a photo of a hacker, might prove to be funny rather than frightening or might cause the public to feel not related to the advertisement.

In order for a campaign to be successful, there are also several pitfalls which need to be avoided. The *first* is not understanding what security awareness really is. *Second*, a compliance awareness program does not necessarily equate to creating the desired behaviours. *Third*, usually there is lack of engaging and appropriate materials. *Fourth*, usually there is no illustration that awareness is a unique discipline. *Fifth*, there is no assessment of the awareness programmes

[37]. *Sixth*, not arranging multiple training exercises but instead focusing on a specific topic or threat does not offer the overall training needed [38].

*Perceived control* and personal handling ability, the sense one has that he/she can drive specific behaviour, has been found to affect the intention of behaviour but also the real behaviour [18][19]. We suggest that a campaign should use simple consistent rules of behaviour that people can follow. This way, their perception of control will lead to better acceptance of the suggested behaviour.

*Cultural differences* in risk perceptions can also influence the maintenance of a particular way of life. Moreover, even when people are willing to change their behaviour, the process of learning a new behaviour needs to be supported [22][23]. We suggest that cultural differences should be taken into consideration while planning a cyber security awareness campaign.

Measuring the effectiveness of information security awareness efforts for the public though, can be a very complicated process. Metrics such as the number of phishing e-mails opened or number of accesses to unauthorised pages are difficult to measure in a larger scale. This is why, defined large scale metrics are needed, to help security-awareness efforts be evaluated and assessed.

The present paper has thus far, reviewed some of the various personal, social and environmental factors influencing online behaviour change as it relates to cyber security. Also, we have tried to identify the factors which can lead to a cyber security awareness campaign's success or failure.

# 6 Case studies

This section will present existing awareness campaigns on cyber security in the UK and in Africa. These two countries were selected in an effort to explore possible core cultural differences reflected in awareness efforts. The two countries differ not only regarding cultural characteristics, but also in the amount of investment being spent on influencing secure behaviour online.

## *6.1 Cyber security awareness campaigns in the UK*

There are various awareness efforts in UK aiming to improve online security for businesses and the public. Below, we present two of the most popular of these.

A) *The GetSafeOnline Campaign* [39] is a jointly-funded initiative between several government departments and the private sector, and focuses on users at home and in businesses. The positive message of ''*Get safe online*'' itself is an intriguing one, and at its core, emphasises to individuals that they have the responsibility for getting safe online. The campaign offers a comprehensive repository of information on threats and how-to advice on protecting oneself and one's enterprise. The charge, however, is on individuals to make use of this information and properly apply it to their context.

B) *The Cyber Streetwise Campaign* [40] also concentrates on users at home and in businesses. The new Home Office Cyber Streetwise site advises businesses to adopt five basic measures to boost their security. These include, using strong, memorable passwords, installing antivirus software on all work devices, checking privacy settings on social media, checking the security of online retailers before loading card details, and patching systems as soon as updates are available. This is a campaign which tries to cause a behavioural change by providing tips and advice on how to improve online security. The campaign uses a positive message method to influence the behaviour of users, ''*In short, the weakest links in the cyber security chain are you and me*''.

## 6.2 Cyber security awareness campaigns in Africa

A) *The ISC Africa* [41] is a coordinated, industry and community-wide effort to inform and educate Africa's citizens on safe and responsible use of computers and the Internet, so that the inherent risks can be minimised and consumer trust can be increased. The campaign uses a positive message method to influence the behaviour of users in a more collectivist approach ''*Working together to ensure a safe online environment for all*'. Here, we can see an obvious difference to the messages used in awareness campaign in the UK, that is, the cyber security-awareness efforts in Africa have been aligned to the cultural aspects of that society.

B) *Parents' Corner Campaign* [42] is intended to co-ordinate the work done by government, industry and civil society. Its objectives are to protect children, empower parents, educate children and create partnerships and collaboration amongst concerned stakeholders. Parents' Corner tips for a safer Internet include: ''*People aren't always who they say they are, Think before you post, Just as they would in real life - friends must protect friends*''. Once again, one of the main messages refer to users protecting users in terms of their relationships and social group memberships.

## 6.3 Comparing cyber security awareness campaigns in the UK and Africa

In our effort to investigate potential differences in cyber security-awareness campaigns, in different cultural contexts, we considered existing campaigns in the UK and Africa.
We have to state that there are a large number of existing national campaigns in the UK, but we selected two of the most popular of these. On the contrary, in Africa the number of existing awareness campaigns is limited. This difference could indicate lack of resources, or lack of current emphasis on cyber security in Africa. Moreover, it could even indicate that Africa has a more organised and coherent security-awareness plan, with a small number of targeted and coordinated campaigns.

As previously discussed, messages and advertisements are usually preferred when they match a cultural theme of the message recipient [23]. While reviewing the main messages used by campaigns in the UK, it became clear that most of them refer to the individual [25]. For example The *GetSafeOnline Campaign uses the message ''Get safe online''* by emphasising to individuals and their responsibility for getting safe online. On the contrary, the messages used by campaigns in Africa, refer to users in terms of their relationships and social group memberships, as well as the need to fulfil duties and obligations (Parents' Corner Campaign includes a message saying: *Just as they would in real life - friends must protect friends*).

The cultural aspects have been reflected in the awareness campaigns, in both cases, using a more individualist approach in UK and a more collectivist approach in Africa [23][25][27]. It is important to decide the target group of a campaign and try to match a cultural theme of the message recipient but also, match the recipient's cognitive, affective or motivational characteristics with the content of framing of the message [27][26][29].

Usually, most of official awareness-campaign sites include advice which usually comes from security experts and service providers, who monotonically repeat suggestions such as 'use strong passwords'. Such advice pushes responsibility and workload for issues that should be addressed by the service providers and product vendors onto users. One of the main reasons why users do not behave optimally is that security systems and policies are often poorly designed [9]. There is a need to move from awareness to tangible behaviours.

Another important aspect is that most of the official awareness-campaign sites in UK and Africa do not offer the possibility to users to call a help-line, not only to report cybercrime but also to receive help. Less skilled users could find this feature useful.

## 7 Conclusions

This paper presents a review of current literature based on the psychological theories of awareness and behaviour in the area of cyber security, and considers them to gain insight into the reasons why security-awareness campaigns often fail.

Simple transfer of knowledge about good practices in security is far from enough [6]. Knowledge and awareness is a prerequisite to change behaviour but not necessarily sufficient, and this is why it has to be implemented in conjunction with other influencing strategies. It is very important to embed positive cyber security behaviours, which can result to thinking becoming a habit, and a part of an organisation's cyber security culture. One of the main reasons why users do not behave optimally is that security systems and policies are poorly designed – this has been presented time and time again throughout research [9].

Behaviour change in a cyber security context could possibly be measured through risk reduction, but not through what people know, what they ignore or what they do not know. Answering questions correctly does not mean that the individual is motivated to behave according to the knowledge gained during an awareness programme. A campaign should use simple consistent rules of behaviour that people can follow. This way, people's perception of control will lead to better acceptance of the suggested behaviour [18][19][20].

Based on our review on the literature and analysis of several successful and unsuccessful security-awareness campaigns, we suggest that the following factors can be extremely helpful at enhancing the effectiveness of current and future campaigns: (1) security awareness has to be professionally prepared and organised in order to work; (2) invoking fear in people is not an effective tactic, since it could scare people who can least afford to take risks [30]; (3) security education has to be more than providing information to users – it needs to be targeted, actionable, doable and provide feedback; (4) once people are willing to change, training and continuous feedback is needed to sustain them through the change period; (5) emphasis is necessary on different cultural contexts and characteristics when creating cyber security-awareness campaigns [35].

In future work, we will aim to conduct a more substantial evaluation of several cyber security-awareness campaigns around the world, especially in North America and Asia, to examine the extent to which they have implemented the factors mentioned above and their levels of campaign success.